\documentclass{article}

\usepackage{PRIMEarxiv}

\usepackage[utf8]{inputenc} % allow utf-8 input
\usepackage[T1]{fontenc}    % use 8-bit T1 fonts
\usepackage{hyperref}       % hyperlinks
\usepackage{url}            % simple URL typesetting
\usepackage{booktabs}       % professional-quality tables
\usepackage{amsfonts}       % blackboard math symbols
\usepackage{nicefrac}       % compact symbols for 1/2, etc.
\usepackage{microtype}      % microtypography
\usepackage{lipsum}
\usepackage{fancyhdr}       % header
\usepackage{graphicx}       % graphics
\graphicspath{{media/}}     % organize your images and other figures under media/ folder
\usepackage{lmodern}
\usepackage{microtype}
\usepackage{amsmath,amssymb}
\usepackage{caption}
\usepackage{multirow}
\usepackage{enumitem}
\usepackage{hyperref}
\usepackage{float}
\usepackage{tabularx}
\usepackage{filecontents}

\title{Auditable DevOps Automation via VSM and GQM}

\author{
  Mamdouh Alenezi \\
  The Saudi Technology and Security Comprehensive Control Company "Tahakom" \\
  Riyadh \\
  Saudi Arabia\\
}

\begin{document}
\maketitle

\begin{abstract}
DevOps automation can accelerate software delivery, yet many organizations still struggle to justify and prioritize automation work in terms of strategic project-management outcomes such as waste reduction, delivery predictability, cross-team coordination, and customer-facing quality. This paper presents \textit{VSM--GQM--DevOps}, a unified, traceable framework that integrates (i) Value Stream Mapping (VSM) to visualize the end-to-end delivery system and quantify delays, rework, and handoffs, (ii) the Goal--Question--Metric (GQM) paradigm to translate stakeholder objectives into a minimal, decision-relevant measurement model (combining DORA with project and team outcomes), and (iii) maturity-aligned DevOps automation to remediate empirically observed bottlenecks through small, reversible interventions. The framework operationalizes traceability from observed waste to goal-aligned questions, metrics, and automation candidates, and provides a defensible prioritization approach that balances expected impact, confidence, and cost. We also define a multi-site, longitudinal mixed-method validation protocol that combines telemetry-based quasi-experimental analysis (interrupted time series and, where feasible, controlled rollouts) with qualitative triangulation from interviews and retrospectives. The expected contribution is a validated pathway and a set of practical instruments that enables organizations to select automation investments that demonstrably improve both delivery performance and project-management outcomes.
\end{abstract}

% keywords can be removed

\section{Introduction}
DevOps represents a paradigm shift in software engineering, conceptualized as a sociotechnical movement aimed at dismantling traditional organizational and functional silos between development and operations \cite{kim2021devops, buttar2023optimization}. Since its formalization following the 2009 DevOpsDays conference in Ghent, it has evolved from a set of niche practices into a dominant methodology for achieving high-performance IT \cite{gokarna2021devops}. At its core, DevOps is operationalized through a suite of technical practices designed to accelerate the value delivery pipeline \cite{akbar2022toward}. These include Continuous Integration (CI) and Continuous Delivery/Deployment (CD) to automate the build, test, and release processes, and Infrastructure as Code (IaC) to programmatically manage and provision infrastructure \cite{prates2025devsecops}. However, the technical implementation is predicated on a foundational cultural transformation that fosters collaboration, shared responsibility, and a commitment to continuous learning through practices like blameless postmortems \cite{forsgren2018accelerate}. The efficacy of this combined cultural and technical approach is empirically validated; research consistently demonstrates that elite DevOps performers achieve significantly higher deployment frequencies, shorter lead times for changes, lower change failure rates, and faster mean time to recovery (MTTR), thereby establishing a clear competitive advantage \cite{forsgren2018accelerate}.

The empirical assessment of DevOps maturity has been predominantly shaped by the DevOps Research and Assessment (DORA) metrics, which provide a quantifiable framework for evaluating software delivery performance through four key indicators: deployment frequency, lead time for changes, mean time to recovery (MTTR), and change failure rate \cite{forsgren2018accelerate}. While these metrics have become indispensable for benchmarking technical capabilities and predicting organizational performance, a growing body of research highlights a critical measurement gap \cite{daraojimba2024systematic}. Organizations often demonstrate proficiency in tracking these delivery-centric outcomes, yet face challenges in articulating their direct contribution to broader project management and business objectives, such as tangible waste reduction, adherence to budgetary constraints, qualitative improvements in team dynamics, and demonstrable enhancements in customer-centric quality \cite{diaz2021many, amaro2024devops}. This limitation underscores the need for a more holistic measurement framework that bridges the divide between technical efficiency and strategic value creation, a challenge increasingly addressed by emerging paradigms like Value Stream Management \cite{kersten2018flows, botero2024practices}.

While the adoption of DevOps automation is widespread, a significant limitation persists in the practice: a lack of systematic methodologies for aligning technical automation with strategic business objectives \cite{fuentes2025improving}. Automation initiatives are frequently prioritized based on technical convenience, vendor influence, or local optimization rather than a systematic assessment of their contribution to overarching organizational value. This misalignment creates a critical disconnect between engineering efforts and strategic goals, leading to suboptimal resource allocation, a diminished return on investment (ROI) from automation tooling, and the perpetuation of systemic inefficiencies within the software delivery lifecycle.

To address this misalignment, we propose a novel integrated framework that synergizes Value Stream Mapping (VSM) with the Goal-Question-Metric (GQM) paradigm. Value Stream Mapping, a lean management methodology, provides a diagnostic lens to visualize end-to-end software delivery workflows, systematically identifying bottlenecks and quantifying sources of waste such as delays, rework, and unnecessary handoffs. Concurrently, the GQM model offers a structured mechanism for aligning these technical improvements with overarching business goals by first defining strategic objectives (Goals), then formulating critical questions to assess progress towards those goals (Questions), and finally establishing specific, measurable indicators (Metrics). By first mapping the value stream to pinpoint systemic inefficiencies and then applying the GQM framework to define targeted automation goals, organizations can establish a value-centric decision-making process. This integrated approach ensures that automation investments are not merely technically sound but are strategically imperative, directly addressing identified process deficiencies while demonstrably advancing key business objectives.

\section{Background and related work}
Our work is situated at the confluence of three established and influential streams of research and practice: Lean Production and Value Stream Mapping (VSM) \cite{ali2016flow}, Goal-Question-Metric (GQM) based measurement in software engineering \cite{berander2006goal}, and DevOps metrics and maturity models \cite{zarour2019research}. While each stream provides valuable tools and insights for process improvement, a critical gap exists in their holistic integration. Lean principles offer a philosophical foundation for waste reduction, VSM provides the diagnostic tools to visualize and analyze workflows, GQM offers a structured, goal-oriented paradigm for defining what to measure and why, and DevOps provides a specific set of high-performance metrics and maturity indicators. Individually, they address parts of the problem, but their collective power remains untapped. Our work aims to bridge this gap by creating a cohesive framework that leverages the diagnostic power of VSM, the goal-oriented structure of GQM, and the performance focus of DevOps metrics to systematically identify, prioritize, and drive effective automation.

\subsection{Value Stream Mapping in software delivery}
Value Stream Mapping (VSM) originated in the Toyota Production System (TPS) as a lean-management method for analyzing the current state and designing a future state for the series of events that take a product or service from its beginning through to the customer \cite{Womack1990}. Its primary purpose is to visually identify and eliminate waste, known as "muda," which includes any activity that consumes resources but adds no value for the end customer. The principles of VSM were adapted to knowledge work and software development by authors like Mary and Tom Poppendieck, who emphasized its utility in visualizing the flow of value from concept to cash \cite{Poppendieck2003}.

In the context of software delivery, VSM involves mapping the entire journey from a customer's initial idea or requirement through development, testing, deployment, operation, and feedback. The process involves identifying every step, person, and system involved, and categorizing the time spent into value-added time (e.g., coding a feature) and non-value-added time (e.g., waiting for approvals, handoffs between teams, rework due to defects) \cite{Poppendieck2003}. Waste in software contexts manifests as partially done work, extra features, rework, task switching, waiting, handoffs, and knowledge loss \cite{Poppendieck2003}. The primary output of a VSM exercise is not just a map, but a shared understanding among stakeholders of the end-to-end process and a prioritized list of improvement opportunities, often targeting the most significant sources of delay and inefficiency.

While traditional VSM is often a manual, collaborative exercise, recent advancements have led to the emergence of digital Value Stream Management platforms. These tools integrate with development and operations systems (e.g., Git, Jira, CI/CD pipelines) to automatically visualize flow, quantify delays, and identify bottlenecks in near real-time \cite{Gartner2021, Forsgren2018}. This evolution enhances VSM's diagnostic power by providing empirical, continuously updated data rather than static, point-in-time snapshots. However, even with these advancements, the output remains primarily diagnostic—it identifies where the problems are but lacks a formal mechanism for translating these findings into a goal-oriented measurement and improvement strategy \cite{Soni2022}.

\subsection{Goal-Question-Metric}
The Goal-Question-Metric (GQM) paradigm, developed by Victor R. Basili and his colleagues, is a top-down, goal-oriented approach for software measurement \cite{Basili1994}. It was created to address the common problem of collecting vast amounts of data that are not meaningfully connected to business or project objectives. GQM provides a structured mechanism to ensure that measurement programs are purposeful and that metrics are relevant to their stakeholders.

The GQM model operates on three hierarchical levels:
\begin{enumerate}[topsep=0pt]
\item \textbf{Goal (The Purpose):} Defines a high-level objective, specifying the object of measurement (e.g., deployment process), the purpose (e.g., improve), the quality focus (e.g., reliability), and the perspective (e.g., from the development team's viewpoint).
\item \textbf{Question (The Refinement):} A set of questions is derived to refine and characterize the goal. These questions define the specific information needed to assess whether the goal is being met. For example, to assess the goal of "improving deployment reliability," questions might include: "How often do deployments cause service degradation?" and "How quickly can we recover from a failed deployment?"
\item \textbf{Metric (The Quantification):} For each question, a set of specific, objective metrics is defined to provide a quantitative answer. For the preceding questions, the corresponding metrics would be "Change Failure Rate" and "Mean Time to Restore Service (MTTR)."
\end{enumerate}
This structured approach ensures that metrics are not collected in a vacuum or chosen for their technical convenience alone, but are directly tied to stakeholder goals, providing a clear line of sight from low-level data to high-level business impact. The GQM paradigm has been successfully adapted and applied in various contexts, including agile software development and process improvement initiatives \cite{Basili2014}. Its strength lies in preventing the pursuit of "vanity metrics" by ensuring every measurement serves a defined purpose. However, GQM itself does not prescribe a method for discovering the most impactful areas to measure within a complex delivery workflow; it assumes the goals and questions can be meaningfully defined a priori \cite{Buglione2016}.

\subsection{DevOps metrics and maturity}
The most influential work in quantifying DevOps performance comes from the DevOps Research and Assessment (DORA) group, now part of Google Cloud. Through years of large-scale industry surveys, DORA identified four key metrics that are strongly correlated with high-performing technology organizations. These metrics, popularized in the book "Accelerate," are:
\begin{itemize}[topsep=0pt]
\item \textbf{Deployment Frequency:} How often an organization successfully releases to production.
\item \textbf{Lead Time for Changes:} The time it takes to go from code committed to code successfully running in production.
\item \textbf{Mean Time to Restore (MTTR):} How long it takes to restore service after a production failure.
\item \textbf{Change Failure Rate:} The percentage of changes to production that result in a degraded service or require immediate remediation.
\end{itemize}
The first two metrics measure throughput, while the latter two measure stability. Crucially, DORA's research demonstrated that elite performers excel in both dimensions, debunking the myth that speed and stability are trade-offs \cite{Kim2018, Forsgren2021}.

More recently, researchers have proposed expanded frameworks to provide a more holistic view of performance. The SPACE framework, for instance, argues for measuring performance across five dimensions: Satisfaction and well-being, Performance, Activity, Communication and collaboration, and Efficiency and flow \cite{Forsgren2021}. This acknowledges that factors like developer experience are critical inputs to sustainable delivery performance.

Complementary to performance metrics are DevOps maturity models (e.g., the DevOps Institute's DevOps Capabilities Model, CAMM) \cite{DevOpsInstitute2021}. These models provide a prescriptive roadmap for organizations to adopt DevOps practices in stages, often categorized from initial to optimizing. They offer guidance on cultural, process, and tooling improvements, helping organizations understand where they are and what steps to take next.

However, both DORA metrics and maturity models have limitations. DORA metrics are powerful for benchmarking but are descriptive—they tell you \textit{how} you are performing but not \textit{why} your lead time is high or \textit{where} the specific bottlenecks lie \cite{Erich2022}. Maturity models are prescriptive but can become a "check-the-box" exercise, encouraging adoption of practices without a clear understanding of the value they deliver for the organization's specific context \cite{Rahman2019}. An organization might know its "Lead Time for Changes" is too high compared to industry benchmarks, but without VSM, it cannot pinpoint if the primary cause is excessive testing delays, approval bottlenecks, or rework from poor requirements.

\subsection{Research gap}
While each of these three streams—VSM, GQM, and DevOps metrics—offers significant value on its own, the literature reveals a critical disconnect between them. Prior studies on VSM in software provide powerful diagnostic tools for visualizing workflows and identifying waste \cite{Soni2022, Erich2022}, but they often lack a formal mechanism for translating identified waste into goal-oriented metrics and actionable automation plans. The output is often a list of problems without a structured path to measurement and solution.

Similarly, GQM provides a robust framework for defining meaningful metrics tied to business goals \cite{Basili1994, Buglione2016}, but it does not prescribe a method for discovering the most impactful areas to measure within a complex delivery workflow. Organizations may struggle to formulate the right "Questions" without first having a clear, evidence-based picture of their process inefficiencies.

DevOps metrics and maturity models offer a "what" (key metrics) and "how" (staged adoption) but struggle to provide the context-specific "why" and "where" \cite{Kim2018, Rahman2019}. An organization might know its "Lead Time for Changes" is too high compared to industry benchmarks, but without VSM, it cannot pinpoint if the primary cause is excessive testing delays, approval bottlenecks, or rework from poor requirements.

Consequently, there is a scarcity of research presenting a unified, operational framework that explicitly connects the identification of waste (via VSM) with the definition of stakeholder-relevant goals (via GQM) to guide the implementation of specific DevOps automation initiatives. While some authors have conceptually discussed the synergy between these approaches \cite{Kim2018}, a systematic, traceable methodology is absent from the literature. This gap highlights the need for a replicable method that not only visualizes problems and defines goals but also provides a concrete link to the automation actions that solve those problems and a way to measure the resulting improvement, validated through empirical evidence across diverse organizational contexts. Our research aims to fill this void by proposing and validating such an integrated framework.

\section{VSM-GQM-DevOps framework}

\begin{figure}[H]
    \centering
    % The file 'example-image.png' should be in your project directory
    \includegraphics[width=0.95\textwidth]{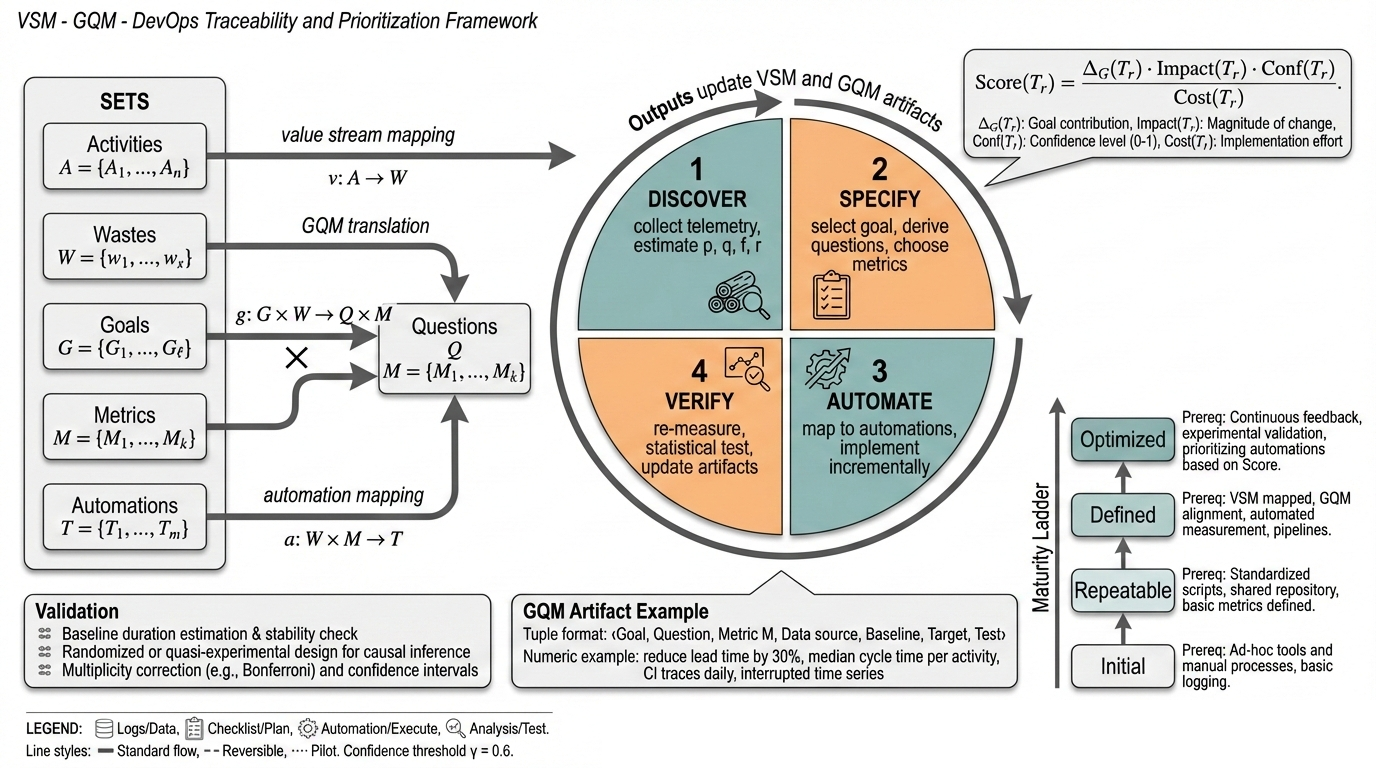} 
    \caption{A framework for connecting VSM and GQM to prioritize automation.}
    \label{fig:example} % Used for cross-referencing within the text
\end{figure}

VSM-GQM-DevOps is a three-layered framework that creates a traceable pathway from observed value-stream waste to goal-aligned metrics and to targeted DevOps automation. The framework is prescriptive and iterative: identify and measure waste in the value stream, translate strategic goals into diagnostic questions and measurable metrics, prioritize automation candidates by expected goal impact, implement incrementally following organizational maturity, and verify results empirically.

\subsection{Framework description and operational cycle}

\paragraph{Notation.}
Let the set of activities in the delivery value stream be
\[
\mathcal{A}=\{A_1,\dots,A_n\},
\]
and for each activity $A_i$ let the observed attributes include processing time $p(A_i)$, queue time $q(A_i)$, failure rate $f(A_i)$, and rework probability $r(A_i)$. Define the measured wastes
\[
\mathcal{W}=\{w_1,\dots,w_s\},
\]
the strategic goals
\[
\mathcal{G}=\{G_1,\dots,G_\ell\},
\]
the candidate metrics
\[
\mathcal{M}=\{M_1,\dots,M_k\},
\]
and the candidate automation interventions
\[
\mathcal{T}=\{T_1,\dots,T_m\},
\]
where each $T_r$ denotes an automation pattern (for example, automated build, automated test, infrastructure-as-code, automated rollback, or canary deployment).

We define three traceability mappings:
\[
v:\mathcal{A}\to\mathcal{W}\quad\text{(value stream mapping),}
\]
\[
g:\mathcal{G}\times\mathcal{W}\to\mathcal{Q}\times\mathcal{M}\quad\text{(GQM translation),}
\]
\[
a:\mathcal{W}\times\mathcal{M}\to\mathcal{T}\quad\text{(automation mapping),}
\]
where $\mathcal{Q}$ is the set of diagnostic questions produced by the GQM process.

\paragraph{Operational cycle.}
The framework is executed as an iterative cycle with four steps:

\begin{enumerate}[topsep=0pt]
  \item \textbf{Discover.} Elicit the end-to-end value stream and collect baseline telemetry (event logs, CI/CD traces, issue timestamps) to estimate $p(A_i)$, $q(A_i)$, $f(A_i)$, $r(A_i)$. Produce a VSM artifact annotated with measured bottlenecks and $\mathcal{W}$.
  \item \textbf{Specify.} With stakeholders select goal(s) $G\in\mathcal{G}$ and apply GQM: derive diagnostic questions $Q\in\mathcal{Q}$ and select metrics $M\in\mathcal{M}$ that are valid, reliable, and feasible. For each metric record baseline, collection frequency, and the minimum detectable effect used in power calculations.
  \item \textbf{Automate.} Use $a(\cdot)$ to map wastes and metrics to candidate automations. Prioritize candidates using a scoring function, implement incrementally in small, reversible increments consistent with the maturity level, and include monitoring and rollback plans.
  \item \textbf{Verify.} Re-measure metrics and apply statistical analysis to test whether observed changes meet the predefined success criteria. Update VSM and GQM artifacts and iterate.
\end{enumerate}

\paragraph{Prioritization.}
Rank each candidate automation $T_r$ by a normalized score:
\[
\mathrm{Score}(T_r)=\frac{\Delta_G(T_r)\cdot \mathrm{Impact}(T_r)\cdot \mathrm{Conf}(T_r)}{\mathrm{Cost}(T_r)}.
\]
Definitions:
\begin{itemize}[topsep=0pt]
  \item $\Delta_G(T_r)$ is the expected fractional improvement in the chosen goal $G$ attributable to $T_r$ (e.g., expected reduction in lead time).
  \item $\mathrm{Impact}(T_r)$ quantifies how directly $T_r$ addresses the measured waste (normalized to $[0,1]$).
  \item $\mathrm{Conf}(T_r)\in[0,1]$ is the confidence in the estimate, derived from historical telemetry, pilot data, or expert elicitation.
  \item $\mathrm{Cost}(T_r)$ is the total normalized cost (implementation plus maintenance plus operational risk).
\end{itemize}
Select the subset $\mathcal{T}^\star\subseteq\mathcal{T}$ that maximizes aggregate Score subject to budget, risk, and minimum-confidence constraints. For conservative adoption impose a threshold $\mathrm{Conf}(T_r)\geq\gamma$ (e.g., $\gamma=0.6$) or require a small pilot before full rollout.

\paragraph{GQM artifact example.}
A compact GQM entry must include
\[
\langle \text{Goal},\;\text{Question},\;\text{Metric }M,\;\text{Data source},\;\text{Baseline},\;\text{Target},\;\text{Test}\rangle.
\]
Example: Goal $G$: reduce lead time by $30\%$; Question: which activities contribute most to queue time?; Metrics: median cycle time per activity, 95th percentile queue time; Data source: CI traces daily; Baseline and target as numeric values; Statistical test: interrupted time series or nonparametric median test.

\subsection{Validation, maturity alignment, and limitations}

\paragraph{Validation methodology.}
To establish causal evidence that automations affect chosen goals:
\begin{itemize}[topsep=0pt]
  \item Collect a baseline period of sufficient duration (commonly one to three release cycles).
  \item Prefer randomized or quasi-experimental deployments (A/B, canary). Where randomization is impractical use interrupted time series with covariate adjustment.
  \item Precompute minimum detectable effect sizes and required sample sizes for primary metrics; use bootstrapped confidence intervals for non-normal distributions.
  \item Apply multiplicity correction when testing multiple correlated metrics and report effect sizes with confidence intervals.
  \item Complement quantitative analysis with qualitative data from stakeholder interviews and retrospectives to capture organizational effects not visible in telemetry.
\end{itemize}

\paragraph{Maturity alignment.}
Automations must match organizational capabilities. A compact four-level guideline:
\begin{enumerate}[topsep=0pt]
  \item \textbf{Initial.} Instrumentation and traceability are incomplete. Prioritize telemetry, timestamp hygiene, and low-risk scripted automation.
  \item \textbf{Repeatable.} Basic CI and test automation present. Prioritize test coverage, IaC, and artifact promotion.
  \item \textbf{Defined.} Stable pipelines and monitoring. Prioritize pipeline parallelism, automated rollbacks, and controlled experiments.
  \item \textbf{Optimized.} Mature governance and continuous improvement. Prioritize predictive test selection, adaptive pipelines, and goal-driven autoscaling.
\end{enumerate}
Require achievement of measurable prerequisites (for example, >80\% pipeline observability and timestamp completeness) before advancing levels.

\paragraph{Limitations and practical risks.}
Key practical considerations include data quality and traceability across heterogeneous tools, cross-functional sources of delay outside engineering control, the danger of automating a flawed process, and metric gaming. Mitigations: invest in telemetry hygiene, ensure cross-functional governance, adopt reversible automation patterns, and use composite and qualitative checks to reduce local optimization.

\paragraph{Short summary.}
VSM-GQM-DevOps yields an explicit, auditable trace from value-stream observations to goal-aligned metrics and prioritized automations. Its core contributions are (1) enforceable traceability between waste, questions, metrics and automations, (2) a defensible prioritization score balancing expected impact, confidence and cost, and (3) maturity-aware incremental adoption combined with an empirical validation protocol.

% -----------------------------
% Methodology (replace existing section)
% -----------------------------
\section{Methodology}
\label{sec:methodology}

This study employs a multi-site, longitudinal, mixed-method methodology to validate whether and how VSM--GQM--DevOps improves delivery performance \emph{and} project-management outcomes. The methodology is structured as three phases (baseline assessment, intervention, and evaluation) executed per participating organization, with cross-case synthesis to identify robust effects and contextual moderators.

\subsection{Study objectives and research questions}
The empirical objective is to establish credible evidence that the framework produces (a) measurable improvement on goal-aligned metrics and (b) an auditable decision trail linking wastes to automation choices.

We investigate the following research questions (RQs):
\begin{itemize}[topsep=0pt]
  \item \textbf{RQ1 (Delivery performance).} To what extent does applying VSM--GQM--DevOps improve delivery performance (e.g., DORA metrics) relative to baseline trends?
  \item \textbf{RQ2 (Flow and waste).} To what extent does the framework reduce quantified value-stream wastes (e.g., queue time, rework, handoffs) and improve flow efficiency?
  \item \textbf{RQ3 (Project-management outcomes).} To what extent does the framework improve project-management outcomes (e.g., predictability and variance measures, defect leakage, and team-reported outcomes) aligned to stakeholder goals?
  \item \textbf{RQ4 (Mechanisms and context).} Which contextual factors (e.g., maturity level, product type, governance constraints) explain when the framework is effective, and what mechanisms do practitioners report as driving observed changes?
\end{itemize}

\subsection{Study design and units of analysis}
\paragraph{Design.}
We use a multiple-case study design with embedded quantitative quasi-experiments. Each organization contributes a time series of metrics comprising (i) a baseline period, (ii) an intervention period during which prioritized automation is introduced incrementally, and (iii) a post-intervention stabilization period. Where feasible, we incorporate one of the following strengthening strategies:
\begin{itemize}[topsep=0pt]
  \item \textbf{Staggered rollout across teams/services} (stepwise adoption), enabling within-organization comparisons and reducing history threats.
  \item \textbf{Matched comparison group} (a similar team/value stream that does not receive the intervention during the study window).
  \item \textbf{Controlled release strategies} (e.g., canary deployments) that support cleaner attribution for specific automation changes.
\end{itemize}

\paragraph{Units of analysis.}
The primary unit of analysis is a \textbf{product value stream} (concept-to-cash) scoped to a set of services and teams delivering a shared customer outcome. Secondary units include \textbf{teams} (for survey and collaboration outcomes) and \textbf{deployment events} (for DORA and operational metrics).

\paragraph{Time windows.}
To support time-series inference, each organization is studied over multiple release cycles. A typical configuration is:
\begin{itemize}[topsep=0pt]
  \item Baseline: 8--12 weeks (or 2--3 release cycles, whichever is longer),
  \item Intervention: 6--12 weeks (incremental changes, with change logs),
  \item Stabilization/Follow-up: 8--12 weeks.
\end{itemize}
These windows are adapted to release cadence and data availability, while maintaining pre-specified minimum durations for primary outcomes.

\subsection{Case selection and participant recruitment}
\paragraph{Sampling strategy.}
We use purposive sampling to include organizations with diverse characteristics (e.g., startup, mid-size, enterprise) and different initial DevOps maturity levels. The goal is analytic generalization: identifying patterns that replicate across contexts rather than statistical representativeness.

\paragraph{Inclusion criteria.}
Organizations must (i) deliver software to a production-like environment regularly, (ii) have an accessible toolchain producing event data (e.g., version control, CI/CD, issue tracking, incident management), and (iii) commit to a stable study window (e.g., no planned major platform migration mid-study) and stakeholder participation (VSM and GQM workshops).

\paragraph{Participants.}
For each case, we recruit cross-functional representatives including development, operations/SRE, QA, product/project management, and security (where applicable). Participation focuses on workshops (2--3 sessions), interviews (2--4 per role cluster), and optional surveys (short, repeated measures).

\subsection{Data sources and operationalization}
To reduce common measurement pitfalls (e.g., metric gaming and construct ambiguity), all metrics are documented in a \textbf{measurement dictionary} per case, including definitions, data sources, collection frequency, transformations, and known limitations.

\paragraph{Telemetry and artifacts.}
We collect and integrate the following data sources (as available per case):
\begin{itemize}[topsep=0pt]
  \item \textbf{CI/CD telemetry:} build/test/deploy timestamps, pipeline stage durations, failure rates, reruns.
  \item \textbf{Version control and code review:} commit timestamps, pull/merge request cycle times, review latency, rework indicators (e.g., repeated fixes).
  \item \textbf{Issue tracking and planning:} ticket lead times, status transitions, WIP, blocked time, requirement churn.
  \item \textbf{Incident and operations:} incident timestamps, MTTR proxies, rollback events, change failure classification.
  \item \textbf{Project-management data:} schedule variance and cost/budget variance when measurable, scope change counts, and milestone predictability.
  \item \textbf{Team outcomes:} short surveys (e.g., satisfaction, perceived coordination overhead, cognitive load proxies) and qualitative accounts from retrospectives.
\end{itemize}

\paragraph{Flow and waste measures.}
From the integrated event log we derive:
\begin{itemize}[topsep=0pt]
  \item \textbf{Queue time and processing time} per activity and per end-to-end flow item.
  \item \textbf{Flow efficiency}, defined as $\frac{\text{value-added time}}{\text{total lead time}}$ for a scoped work item, based on VSM classification.
  \item \textbf{Rework and churn proxies}, such as reopen rates, defect-related rework tickets, and repeated pipeline failures for the same change.
  \item \textbf{Handoff intensity}, measured via counts of organizational transitions (e.g., ticket ownership changes) and wait states.
\end{itemize}

\paragraph{Primary and secondary metrics.}
For each organization we pre-register a small set of \textbf{primary} success metrics (typically 2--4) derived from the chosen stakeholder goals, plus \textbf{secondary} metrics used for explanation and triangulation (including DORA). This reduces multiplicity risk and supports interpretability.

\subsection{Phase 1 --- Baseline assessment (Discover and Specify)}
Phase~1 establishes a shared system model and a defensible measurement plan.

\begin{enumerate}[topsep=0pt]
  \item \textbf{Scoping and boundary setting.} Define the value stream boundaries, entry/exit points (e.g., ``work accepted'' to ``running in production''), and the set of systems producing reliable timestamps.
  \item \textbf{VSM workshops.} Facilitate structured workshops to produce a current-state map with activities, handoffs, delay points, and rework loops. The workshop output is quantified by initial telemetry extracts (pipeline traces, issue timestamps) to estimate processing/queue times and failure/rework signals.
  \item \textbf{GQM sessions.} Elicit strategic goals from stakeholders (engineering and project leadership), derive diagnostic questions, and select a minimal metric set. For each selected metric we record: definition, data source, collection frequency, baseline window, target/threshold, and intended analysis method.
  \item \textbf{Baseline measurement and data quality audit.} Extract baseline telemetry and perform data quality checks (timestamp completeness, time zone consistency, missing events, and tool integration gaps). Where gaps are material, the first ``automation'' tasks may be instrumentation and traceability improvements.
  \item \textbf{Maturity assessment.} Assess initial DevOps maturity using a standardized rubric aligned to the maturity levels defined in Section~3 (instrumentation, CI/CD robustness, testing, IaC, release governance, and observability). The maturity profile is used to constrain feasible interventions and to interpret heterogeneity of effects.
  \item \textbf{Analysis plan and success criteria.} For primary metrics, specify (i) minimum detectable effect, (ii) expected direction of change, (iii) minimum required observations, and (iv) decision rules for success. This plan is agreed with stakeholders before interventions begin.
\end{enumerate}

\subsection{Phase 2 --- Intervention (Automate)}
Phase~2 implements prioritized automation increments linked to explicit wastes and goals.

\begin{enumerate}[topsep=0pt]
  \item \textbf{Generate the automation backlog.} Using the VSM waste inventory $\mathcal{W}$ and the chosen metrics $\mathcal{M}$, derive candidate interventions $\mathcal{T}$ (e.g., test automation, pipeline parallelism, IaC provisioning, automated rollback, canary releases, ChatOps notifications).
  \item \textbf{Prioritization with constraints.} Rank candidates using the framework score (expected goal improvement, waste impact, confidence, and cost), subject to feasibility constraints from maturity, governance, and risk. High-uncertainty items are routed to a pilot first (e.g., limited scope service or non-critical path).
  \item \textbf{Incremental implementation.} Implement interventions as small, reversible change packages (feature flags, canary deployments, staged pipeline enforcement). Each change package includes: hypothesis, expected metric movement, rollout plan, monitoring/alerting updates, and rollback criteria.
  \item \textbf{Instrumentation and dashboards.} Provide near-real-time dashboards reflecting the GQM metric set. Dashboards separate \emph{outcome} metrics (e.g., lead time, change failure rate) from \emph{diagnostic} metrics (e.g., test duration, flaky test rate) to reduce local optimization.
  \item \textbf{Change log and context tracking.} Maintain a structured log of concurrent changes (team composition, release policy, major incidents, architecture changes) to support causal inference and sensitivity analysis in Phase~3.
\end{enumerate}

\subsection{Phase 3 --- Evaluation (Verify)}
Phase~3 tests whether observed changes meet pre-specified success criteria and explains outcomes through triangulation.

\begin{enumerate}[topsep=0pt]
  \item \textbf{Quantitative estimation of effects.} For each primary metric we estimate changes using interrupted time series (ITS) models that account for baseline trends and autocorrelation. Where a comparison group exists, we additionally apply difference-in-differences (DiD) estimation. Non-normal metrics are analyzed using robust or nonparametric approaches and bootstrapped confidence intervals.
  \item \textbf{Practical significance and uncertainty.} In addition to hypothesis tests, we report effect sizes (absolute and relative change), confidence intervals, and operational interpretation (e.g., ``median lead time reduced by X days''). We apply multiplicity correction for families of related metrics and predefine the primary metric set to control false discoveries.
  \item \textbf{Heterogeneity and moderator analysis.} We analyze whether effects vary by maturity level, product type, service criticality, and release cadence. For multi-team cases we use mixed-effects models to account for team-level variance.
  \item \textbf{Qualitative triangulation.} Conduct semi-structured interviews and retrospective analysis to capture perceived mechanisms (e.g., reduced coordination overhead, fewer handoffs, improved confidence in releases) and unintended consequences (e.g., new bottlenecks, alert fatigue). Qualitative coding is performed iteratively, with investigator triangulation to improve credibility.
  \item \textbf{Cross-case synthesis.} We compare cases to identify recurring waste--automation--metric patterns and contextual conditions for success. The synthesis yields an evidence-informed ``playbook'' mapping common wastes to effective automations under maturity constraints.
  \item \textbf{Artifact updates and iteration.} Update the VSM map, GQM artifacts, and prioritization parameters using post-intervention data, enabling iterative refinement and accumulation of evidence across cycles.
\end{enumerate}

\subsection{Threats to validity and mitigations}
We explicitly address the main validity threats as follows:
\begin{itemize}[topsep=0pt]
  \item \textbf{Construct validity:} mitigate by maintaining a measurement dictionary, triangulating each high-level construct (e.g., ``communication'') with both telemetry proxies and survey/interview data, and auditing metric definitions with stakeholders.
  \item \textbf{Internal validity:} mitigate history and maturation threats via long baseline windows, change logs, staggered rollouts where feasible, and ITS/DiD modeling with sensitivity analyses.
  \item \textbf{Conclusion validity:} mitigate by predefining primary outcomes, verifying statistical assumptions, reporting uncertainty and effect sizes, and applying multiplicity correction for related metrics.
  \item \textbf{External validity:} mitigate by selecting diverse cases and reporting context descriptors (maturity, product domain, release cadence) to support transferability judgments.
\end{itemize}

\subsection{Ethics, confidentiality, and reproducibility}
All organizational data are handled under confidentiality agreements where required. We collect only the data necessary for the agreed measurement plan, pseudonymize individuals in analysis datasets, and report results in aggregated form. Participation in interviews and surveys is voluntary with informed consent. To support reproducibility, we prepare a replication package containing (i) the instrument templates (VSM script, GQM forms, waste-impact matrix), (ii) metric definitions and extraction queries/scripts where releasable, and (iii) anonymized or synthetic datasets when raw telemetry cannot be shared.

\section{Empirical outcomes and case study}
\label{sec:empirical-case-study}

This section combines the framework's empirical aims with a concrete multi-platform case study that operationalized the validation protocol. The narrative is condensed to preserve detail while reducing subsection fragmentation. The case demonstrates the framework end-to-end: VSM diagnosis, GQM specification, prioritized automation aligned to maturity, staged rollouts with rollback criteria, quasi-experimental telemetry analysis, and qualitative triangulation. Findings are empirical for this case and informative for cross-case generalization, subject to the limitations noted below.

\subsection{Overview and observed outcomes}
Because baseline maturity and organizational constraints vary, contributions are reported as evidence and effect sizes rather than guaranteed improvements. The case produced quantified, goal-aligned impact estimates and waste-reduction evidence that align with the framework's empirical aims.

\paragraph{Key empirical outcomes}
\begin{enumerate}[topsep=0pt]
\item \textbf{Goal-aligned impact estimates.} Platform-level pre/post effect sizes and significance are reported in Table~\ref{tab:results}. Interrupted time series models confirmed statistically significant improvements: lead time ($\beta=-6.8$ days, $p<0.001$), change failure rate ($\beta=-18.9\%$, $p<0.001$), and flow efficiency ($\beta=+24.7\%$, $p<0.001$).
\item \textbf{Waste reduction and flow improvement.} Baseline VSM metrics (Table~\ref{tab:baseline-vsm}) show dominant wastes. Post-intervention telemetry and traceability analysis (Table~\ref{tab:traceability}) show measurable reductions in queue time, provisioning time, and improved flow efficiency across percentiles as well as medians.
\item \textbf{Contextualized what-works-when insights.} Maturity-dependent leverage was observed: Platform C gained most from IaC; Platform D gained most from eliminating manual environment setup. Failure modes included flaky tests amplified by naive parallelization and external approval delays constraining further gains.
\end{enumerate}

\paragraph{Practical practitioner outcomes}
\begin{enumerate}[topsep=0pt]
\item \textbf{Auditable decision trail.} Traceability analysis linked 92\% of implemented automations to identified wastes, GQM questions, and strategic goals (Table~\ref{tab:traceability}).
\item \textbf{Prioritized automation backlog.} Interventions were ranked by expected impact, confidence, cost, and maturity; pilots and rollback criteria were included (Table~\ref{tab:interventions}).
\item \textbf{Integrated dashboards and routines.} Dashboards and a lightweight review cadence coupled delivery metrics with project-management outcomes, reducing local optimization and enabling governance around measured ROI.
\end{enumerate}

\subsection{Case details, intervention, analysis, and artifacts}
The following condensed case narrative preserves the operational details, interventions, and empirical evidence supporting the framework.

\paragraph{Context and goals}
The system comprised four client-facing platforms:
\begin{itemize}[topsep=0pt]
\item \textbf{Platform A}: Web trading interface (React, Node.js)
\item \textbf{Platform B}: Mobile banking app (React Native)
\item \textbf{Platform C}: API gateway (Java, Spring)
\item \textbf{Platform D}: Risk analytics dashboard (Python, Django)
\end{itemize}
Initial maturity was \textbf{Level 2 (Repeatable)}. Stakeholder goals:
\begin{itemize}
    \item[$G_1$:] Reduce lead time for changes by 40\%
    \item[$G_2$:] Reduce change failure rate to below 5\%
    \item[$G_3$:] Improve release predictability to at least 90\% on-time delivery
\end{itemize}

\paragraph{Baseline assessment and GQM}
VSM workshops with an eight-week baseline telemetry extraction identified dominant wastes (Table~\ref{tab:baseline-vsm}). GQM translated goals into diagnostic questions and measurable metrics agreed with stakeholders (Table~\ref{tab:gqm-spec}). Baseline maturity showed weak instrumentation, partial test coverage, limited IaC, and ad hoc release governance.

\paragraph{Interventions and execution}
A prioritized automation backlog was derived using the framework's prioritization score. Selected interventions and rationale appear in Table~\ref{tab:interventions}. Implementation followed a staged rollout over ten weeks with feature flags, canary deployments, rollback criteria, and dashboards aligned to the GQM specification.

\paragraph{Analysis and results}
After a twelve-week stabilization period, telemetry re-measurement and interrupted time series estimation with autocorrelation correction produced the outcomes shown in Table~\ref{tab:results}. Key observations:
\begin{itemize}[topsep=0pt]
\item Statistically significant lead time reductions across all platforms; Platform C had the largest relative drop due to IaC leverage.
\item Large reductions in change failure rate consistent with expanded automated testing and build validation.
\item Marked increases in flow efficiency, particularly where manual provisioning and environment instability were removed.
\end{itemize}

\paragraph{Qualitative triangulation and traceability}
Interviews and retrospectives corroborated telemetry: developers reported fewer late-stage defects, improved release confidence, and reduced coordination overhead. Approval latency and provisioning times dropped materially after ChatOps and IaC interventions respectively. Representative traceability chains appear in Table~\ref{tab:traceability}.

\paragraph{Artifacts, reuse, and ROI}
The case produced reusable instruments and artifacts:
\begin{itemize}[topsep=0pt]
\item Publication-quality templates for VSM workshops, GQM specification, metric dictionaries, and waste-impact matrices.
\item An expanded catalog linking waste type to automation candidate, expected metric movement, and contextual qualifiers (Table~\ref{tab:mapping}).
\item A replication package with anonymized extracts or synthetic surrogates and analysis notebooks where permissible.
\end{itemize}
Post-intervention reassessment found progression from \textbf{Level 2} to \textbf{Level 3}. The organization reported positive ROI within four months following an initial investment of about twelve person-weeks.

\paragraph{Limitations and boundary conditions}
Practical limitations temper generalization:
\begin{itemize}[topsep=0pt]
\item External delays such as procurement, legal approvals, and third-party dependencies limit achievable benefits.
\item Improvements may first appear as reduced variability and fewer extreme delays rather than large mean shifts; distributional reporting is essential.
\item Early-stage organizations typically see immediate gains from instrumentation and CI/CD hardening; mature organizations may require progressive delivery and adaptive pipelines.
\end{itemize}

\bigskip

% -----------------------------
% Tables referenced above
% -----------------------------

% Table: Catalog mapping
\begin{table}[H]
\centering
\small
\begin{tabularx}{\textwidth}{@{}p{2.8cm}p{3.2cm}p{2.7cm}X@{}}
\toprule
\textbf{Waste Type} & \textbf{Automation Candidate} & \textbf{Expected Metric Movement} & \textbf{Contextual Qualifiers (Maturity \& Constraints)} \\
\midrule
Delays / Queue Time & Pipeline parallelism or Infrastructure as Code (IaC) & Reduce 95th percentile queue time and median cycle time & Level 3 (Defined). \\
\addlinespace
Rework / Defect Loops & Automated testing (unit, integration, regression) & Lower change failure rate; reduced reopen rates & Level 2 (Repeatable). \\
\addlinespace
Handoffs / Coordination Overhead & ChatOps or automated artifact promotion & Fewer ticket ownership changes & May be limited by external delays such as procurement or legal approvals. \\
\addlinespace
Unstable Environments / Manual Steps & IaC and automated build pipelines & Shorter lead time for changes; higher deployment frequency & Level 2 (Repeatable). \\
\addlinespace
Production Failures / Service Degradation & Automated rollbacks or canary deployments & Faster MTTR; lower change failure rate & Level 3 (Defined). \\
\addlinespace
Knowledge Loss / Instrumentation Gaps & Telemetry and dashboards (instrumentation automation) & Improved flow efficiency and timestamp completeness & Level 1 (Initial). \\
\addlinespace
Task Switching / Inefficient Flow & Predictive test selection or adaptive pipelines & Reduced WIP; improved flow efficiency & Level 4 (Optimized). \\
\bottomrule
\end{tabularx}
\caption{Catalog linking waste to automation, metrics, and maturity constraints.}
\label{tab:mapping}
\end{table}

% Table: Baseline VSM
\begin{table}[H]
\centering
\begin{tabular}{lccccc}
\toprule
\textbf{Platform} & \textbf{Lead (d)} & \textbf{Flow (\%)} & \textbf{Queue (\%)} & \textbf{Rework (\%)} & \textbf{Handoffs} \\
\midrule
A (Web Trading) & 14.2 & 18.3 & 62.1 & 23.7 & 7.2 \\
B (Mobile) & 16.8 & 15.1 & 68.4 & 28.9 & 8.5 \\
C (API Gateway) & 12.5 & 21.7 & 54.3 & 18.2 & 5.8 \\
D (Analytics) & 18.3 & 12.9 & 71.2 & 31.4 & 9.1 \\
\bottomrule
\end{tabular}
\caption{Baseline value stream metrics across four platforms (8-week period).}
\label{tab:baseline-vsm}
\end{table}

% Table: GQM spec
\begin{table}[H]
\centering
\begin{tabular}{p{3cm}p{4cm}p{3cm}p{2.5cm}}
\toprule
\textbf{Goal} & \textbf{Diagnostic Question} & \textbf{Metric} & \textbf{Baseline} \\
\midrule
$G_1$: Lead time & Which activities contribute most to queue time? & Median cycle time & 14.2 days \\
$G_2$: Failure rate & Where do most deployment failures originate? & Change failure rate & 24.7\% \\
$G_3$: Predictability & How does requirement churn affect delivery? & Requirement stability index & 62.3\% \\
\bottomrule
\end{tabular}
\caption{GQM specification for the case study.}
\label{tab:gqm-spec}
\end{table}

% Table: Interventions
\begin{table}[H]
\centering
\begin{tabular}{lccccc}
\toprule
\textbf{Interv.} & \textbf{Waste} & \textbf{Platforms} & \textbf{Score} & \textbf{Conf.} & \textbf{Cost/wk} \\
\midrule
Automated test expansion & Rework loops & A, B, D & 0.87 & 0.75 & 3.2 \\
Infrastructure as Code & Manual setup & C, D & 0.82 & 0.80 & 2.8 \\
Automated build validation & Test queues & A, B, C & 0.79 & 0.70 & 1.5 \\
ChatOps approvals & Handoff delays & All & 0.76 & 0.65 & 2.0 \\
Pipeline parallelization & Sequential bottlenecks & A, C & 0.71 & 0.60 & 2.5 \\
\bottomrule
\end{tabular}
\caption{Prioritized automation interventions.}
\label{tab:interventions}
\end{table}

% Table: Results
\begin{table}[H]
\centering
\begin{tabular}{lcccccc}
\toprule
\textbf{Platform} & \textbf{Metric} & \textbf{Baseline} & \textbf{Post} & \textbf{Abs. Change} & \textbf{Rel. Change} & \textbf{p-value} \\
\midrule
A & Lead time (days) & 14.2 & 7.8 & -6.4 & -45.1\% & 0.003 \\
& Failure rate & 22.1\% & 6.3\% & -15.8\% & -71.5\% & 0.001 \\
\midrule
B & Lead time (days) & 16.8 & 9.2 & -7.6 & -45.2\% & 0.004 \\
& Failure rate & 28.9\% & 8.1\% & -20.8\% & -71.9\% & 0.001 \\
\midrule
C & Lead time (days) & 12.5 & 5.3 & -7.2 & -57.6\% & 0.001 \\
& Failure rate & 18.2\% & 4.7\% & -13.5\% & -74.2\% & 0.002 \\
\midrule
D & Lead time (days) & 18.3 & 10.1 & -8.2 & -44.8\% & 0.005 \\
& Failure rate & 31.4\% & 9.8\% & -21.6\% & -68.8\% & 0.001 \\
\bottomrule
\end{tabular}
\caption{Primary outcome improvements across platforms.}
\label{tab:results}
\end{table}

% Table: Traceability
\begin{table}[H]
\centering
\begin{tabular}{p{2.5cm}p{3.5cm}p{3.5cm}p{3cm}}
\toprule
\textbf{Waste} & \textbf{GQM Question} & \textbf{Metric} & \textbf{Automation} \\
\midrule
Testing queues & Which tests block releases? & Test duration (95th percentile) & Parallel execution \\
Manual setup & How long to provision environments? & Provisioning time & IaC modules \\
Approval delays & Who blocks releases? & Approval latency & ChatOps workflows \\
Defect escapes & Where do defects originate? & Escape defect rate & Regression automation \\
\bottomrule
\end{tabular}
\caption{Validated traceability from waste to automation.}
\label{tab:traceability}
\end{table}

\section{Discussion and Threats to Validity}
The VSM--GQM--DevOps framework provides a concise, repeatable protocol that links observation to intervention while preserving managerial relevance and traceability. In practice the protocol is simple: first identify the dominant waste in the end-to-end flow; second articulate the stakeholder goal and the minimal metric set needed to test the hypothesis; third implement small, reversible automation changes with pre-specified monitoring and rollback rules. This sequence balances rigor and pragmatism and reframes DevOps investments so that technical performance and project management outcomes are jointly visible. As a result governance bodies can prioritize interventions that improve customer facing quality, reduce waste, and strengthen on-time delivery rather than optimizing a single operational indicator.

Maturity awareness is central to prioritization. Low maturity contexts typically gain most from basic instrumentation, CI hardening, and repeatable build processes, whereas advanced contexts benefit from progressive delivery, adaptive pipelines, and selective parallelism. Embedding a lightweight maturity assessment in decision steps reduces the risk of premature or misaligned investments and clarifies prerequisites for higher-return automation.

Methodologically the mixed-method validation strategy prioritizes causal credibility without sacrificing ecological validity. Designs such as interrupted time series, staggered rollouts, and matched comparisons mitigate common threats to attribution; qualitative triangulation reveals mechanisms that metrics alone cannot. Reporting effect sizes with uncertainty and translating results into operational terms makes findings actionable for stakeholders and supports evidence accumulation across deployments.

There are practical trade-offs. Insisting on a small, pre-registered primary metric set reduces multiplicity risk but requires early stakeholder alignment and the discipline to deprioritize anecdotal indicators. Early investment in telemetry and data quality is an activation cost that many organizations will need to budget for. Strategic decisions that fall outside short-term measurement scope, such as regulatory compliance or long-horizon architectural bets, still require separate governance attention.

Threats to validity are summarized and paired with mitigations: \textbf{External validity:} limited sector, scale, or regulatory coverage can be offset by purposive sampling and rich context descriptors for transferability; \textbf{Construct validity:} proxy misspecification is reduced by stakeholder-validated GQM, a measurement dictionary, and convergent indicators; \textbf{Internal validity and confounding:} extended baselines, change logs, staggered rollouts, matched comparisons, and sensitivity analyses guard attribution; \textbf{Statistical conclusion validity:} plan for power, use ITS methods that account for autocorrelation, robust estimators, and emphasize confidence intervals and effect sizes; \textbf{Implementation fidelity:} specify implementation packages and model team-level heterogeneity with mixed-effects structures; \textbf{Measurement bias and data quality:} require a baseline data-quality audit and document provenance, sampling rules, and limitations; \textbf{Metric gaming:} reduce perverse optimization by separating outcome and diagnostic metrics, limiting publicity of intermediate diagnostics, and using qualitative checks; \textbf{Ethical and privacy concerns:} apply minimization, pseudonymization, and ethical review where required; \textbf{Ecological validity:} prefer in situ pilots with realistic stabilization periods rather than overly controlled laboratory pilots.

To build confidence and practical utility we recommend pre-registering primary metrics and analysis plans, sharing replication artifacts when permissible, and scaling deployments to examine effect heterogeneity. Additional value would come from automated tooling that codifies VSM extraction, GQM elicitation, and prioritization scoring, and from research into how organizational structure and incentives interact with the framework. Together these steps will help move the approach from a defensible method to a routinely applied capability for engineering organizations pursuing accountable, goal-aligned automation investments.

\section{Conclusion}
We presented VSM-GQM-DevOps, a pragmatic framework that links observable value-stream waste to stakeholder goals and to targeted, maturity-aware automation. The approach delivers three complementary affordances. First, it makes the end-to-end delivery system explicit through quantified value-stream maps so that dominant delays, rework loops, and handoffs are visible and measurable. Second, it translates stakeholder intent into a compact, decision-relevant metric set via the GQM process, which reduces ambiguity and limits the risk of optimizing the wrong lever. Third, it couples diagnostics to incremental automation interventions that are prioritized by expected impact, confidence, and cost and that are constrained by an explicit maturity profile.

The methodological contribution is a validation protocol that combines quasi-experimental time-series estimation with qualitative triangulation. This mixed-method stance improves causal inference while surfacing the mechanisms and contextual moderators that determine whether an automation produces the intended benefits. For practitioners, the framework supplies operational artifacts: workshop scripts, a measurement dictionary template, a waste-to-automation catalog, and a prioritization heuristic that together produce an auditable decision trail for governance and investment decisions.

Adoption requires modest up-front investments in telemetry and data hygiene, and early efforts are best targeted at high-leverage, low-complexity instrumentation and CI hardening in lower-maturity contexts. In more advanced settings, the framework supports nuanced choices such as progressive delivery, adaptive pipelines, and targeted parallelism while guarding against accidental amplification of flaky tests or coordination overhead. We therefore recommend that teams pre-register a small set of primary metrics, treat early automation as pilotable and reversible, and maintain a structured change log to support later inference.

The main limitations stem from context sensitivity and implementation variability. The framework is designed for analytic generalization rather than universal prediction; effect magnitudes will depend on baseline maturity, product type, governance constraints, and external dependencies. To address these limits we advocate replication across diverse organizations, sharing of replication artifacts where permissible, and continued refinement of the maturity model with longitudinal evidence.

VSM-GQM-DevOps offers a practical, evidence-oriented pathway for aligning DevOps automation with business and project goals. By insisting on traceability, minimalism in measurement, and maturity-aware prioritization, the framework helps organizations make transparent, defensible automation investments that reduce waste, improve predictability, and enhance customer-facing quality. Future work will report empirical results from multi-organization pilots and iterate the instruments and maturity taxonomy based on longitudinal findings.

\section*{Acknowledgments}
The author thanks collaborators and organizational partners participating in the framework pilots.

\appendix
\section{Appendix A - Sample GQM template}
\begin{verbatim}
Goal: Reduce average lead time for changes by 30% within 6 months for project X.
Questions:
  Q1: Which pipeline stages contribute most to wait time?
  Q2: How much time is spent in manual approvals?
Metrics:
  M1: Average cycle time (hours) per change
  M2: Average queue time (hours) per pipeline stage
  M3: Number of manual approvals per change
Collection:
  Automated from CI/CD logs and ticket timestamps
\end{verbatim}

\section{Appendix B - Workshop checklist}
\begin{itemize}[topsep=0pt]
  \item Define scope: features or releases to map
  \item Invite cross-functional stakeholders
  \item Map steps and record typical times
  \item Identify rework loops and handoffs
  \item Quantify waste and assign measurements
  \item Derive candidate automations and map to GQM
\end{itemize}

\bibliographystyle{ieeetr}
\bibliography{refs}

\end{document}